\documentclass{article}
\usepackage{lscape}
\usepackage{amssymb}

\usepackage{epsfig}

\usepackage{graphicx,psfrag,amsmath}

\begin{document}

\def\beq#1\eeq{\begin{equation}#1\end{equation}}
\def\beql#1#2\eeql{\begin{equation}\label{#1}#2\end{equation}}

\def\bea#1\eea{\begin{eqnarray}#1\end{eqnarray}}
\def\beal#1#2\eeal{\begin{eqnarray}\label{#1}#2\end{eqnarray}}

\newcommand{\Z}{{\mathbb Z}}
\newcommand{\N}{{\mathbb N}}
\newcommand{\C}{{\mathbb C}}
\newcommand{\Cs}{{\mathbb C}^{*}}
\newcommand{\R}{{\mathbb R}}
\newcommand{\intT}{\int_{[-\pi,\pi]^2}dt_1dt_2}
\newcommand{\cC}{{\mathcal C}}
\newcommand{\cI}{{\mathcal I}}
\newcommand{\cN}{{\mathcal N}}
\newcommand{\cE}{{\mathcal E}}
\newcommand{\cA}{{\mathcal A}}
\newcommand{\xdT}{\dot{{\bf x}}^T}
\newcommand{\bDe}{{\bf \Delta}}

\def\ket#1{\left| #1\right\rangle }
\def\bra#1{\left\langle #1\right| }
\def\braket#1#2{\left\langle #1\vphantom{#2}
  \right. \kern-2.5pt\left| #2\vphantom{#1}\right\rangle }
\newcommand{\gme}[3]{\bra{#1}#3\ket{#2}}
\newcommand{\ome}[2]{\gme{#1}{#2}{\mathcal{O}}}
\newcommand{\spr}[2]{\braket{#1}{#2}}
\newcommand{\eq}[1]{Eq\,\ref{#1}}
\newcommand{\xp}[1]{e^{#1}}

\def\limfunc#1{\mathop{\rm #1}}
\def\Tr{\limfunc{Tr}}

\def\dr{detector }
\def\drn{detector}
\def\dtn{detection }
\def\dtnn{detection}

\def\pho{photon }
\def\phon{photon}
\def\phos{photons }
\def\phosn{photons}
\def\mmt{measurement }
\def\an{amplitude}
\def\a{amplitude }
\def\co{coherence }
\def\con{coherence}

\def\st{state }
\def\stn{state}
\def\sts{states }
\def\stsn{states}

\def\cow{``collapse of the wavefunction" }
\def\cown{``collapse of the wavefunction"}
\def\de{decoherence }
\def\den{decoherence}
\def\dm{density matrix }
\def\dmn{density matrix}
\def\dms{density matrices }
\def\dmsn{density matrices}

\newcommand{\mop}{\cal O }
\newcommand{\dt}{{d\over dt}}
\def\qm{quantum mechanics }
\def\qms{quantum mechanics }
\def\qml{quantum mechanical }

\def\qmn{quantum mechanics}
\def\mmtn{measurement}
\def\pow{preparation of the wavefunction }

\def\me{ L.~Stodolsky }
\def\T{temperature }
\def\Tn{temperature}
\def\t{time }
\def\tn{time}
\def\wfs{wavefunctions }
\def\wf{wavefunction }
\def\wfn{wavefunction} 
\def\wfsn{wavefunctions}
\def\wvp{wavepacket }
\def\pa{probability amplitude } 
\def\sy{system } 
\def\sys{systems }
\def\syn{system} 
\def\sysn{systems} 
\def\ha{hamiltonian }
\def\han{hamiltonian}
\def\rh{$\rho$ }
\def\rhn{$\rho$}
\def\op{$\cal O$ }
\def\opn{$\cal O$}
\def\yy{energy }
\def\yyn{energy}
\def\yys{energies }
\def\yysn{energies}
\def\pz{$\bf P$ }
\def\pzn{$\bf P$}
\def\pl{particle }
\def\pls{particles }
\def\pln{particle}
\def\plsn{particles}

\def\plz{polarization  }
\def\plzs{polarizations }
\def\plzn{polarization}
\def\plzsn{polarizations}

\def\sctg{scattering }
\def\sctgn{scattering}

\def\prob{probability }
\def\probn{probability}

\def\om{\omega} 

\def\hf{\tfrac{1}{2}}

\def\zz{neutrino }
\def\zzn{neutrino}
\def\zzs{neutrinos }
\def\zzsn{neutrinos}

\def\zn{neutron }
\def\znn{neutron}
\def\zns{neutrons }
\def\znsn{neutrons}

\def\csss{cross section }
\def\csssn{cross section}

\def\jp{$J/\psi$ }
\def\b{$B^o$ }
\def\k{$K^o$ }
\def\d{$D^o$ }
\def\db{$\bar D^o$ }
\def\ks{$K_s$ }
\def\kl{$K_l$ }
\def\bb{$\bar B^o$}
\def\kb{$\bar K^{o^{~}}$}

\def\jpn{J/\psi}
\def\bn{B^o}
\def\kn{K^o}
\def\ksn{K_s}
\def\kln{K_l}
\def\bbn{\bar {B^o }}
\def\kbn{\bar {K^o }}

\def\dn{{D^o }}
\def\dbn{\bar {D^o }}

\def\sm{Standard Model }
\def\smn{Standard Model}

\def\sgx{\sigma_1}
\def\sgy{\sigma_2}
\def\sgz{\sigma_3}

\def\fth{\tfrac{1}{4}}

\def\cv{CP violation }
\def\cvn{CP violation}

\def\dcv{direct CP violation }
\def\dcvn{direct CP violation}

\def\as{`away side'}
\def\ns{`near side'}

\def\co{`collapse' }
\def\con{`collapse'}

\def\dg{${ \Delta  \Gamma}$ }
\def\dgn{{ \Delta \bf \Gamma} }

\title{Refined Applications of the\\``Collapse of the
Wavefunction''
}

\author{
 L. Stodolsky\\
Max-Planck-Institut f\"ur Physik
(Werner-Heisenberg-Institut)\\
F\"ohringer Ring 6, 80805 M\"unchen, Germany}

\maketitle

\begin{abstract}

 In a two-part \sy the \cow of one part  can  put the other
part in a state which  would  be  difficult or  impossible  to
achieve otherwise, in particular one  sensitive to small
effects in the `collapse' interaction.

We present some applications to the very symmeteric and
experimentally accessible situations
 of the decays $\phi(1020)\to \kn\kn$,
$\psi(3770)\to \dn\dn$, or  $\Upsilon(4s)\to \bn\bn$, involving
the internal state of the two-state  \k \d or \b  mesons.  
The \cow occasioned by a decay of one member of the pair (\as)
 fixes the  state vector of that side's  two-state \syn.
Bose-Einstein
statistics then determines the state of the recoiling meson (\ns),
whose evolution can then  be followed further.

  In particular the statistics requirement  dictates
that the `away side' and `near side' internal \wfs  must be
orthogonal at the time of the ``collapse''. Thus a CP violation in
the \as\, decay  implies a
complementary CP impurity on the `near side', which can  be
detected in the further evolution.
 The
\cv so manifested is necessarily {\it direct }\cvn, since neither
the mass matrix
nor  time evolution was involved in the `` collapse''. 

 A parametrization of the \dcv is given and  various manifestations
are presented.  Certain rates or combination
of rates are identified which  are nonzero only if there
is \dcvn. 

  The very
explicit and detailed use made of \cow  makes the procedure
interesting  with respect to the fundamentals of \qmn.  We note an
experimental consistency test for our
treatment of the \cown, which can be carried out by a certain
measurement of partial  decay rates.

\end{abstract}

\section{Introduction}
 The \cown, where a "measurement" suddenly fixes the state of a
\qml \syn,
is one of the longest discussed and most difficult chapters in
\qmn. This is especially true when the `collapse' is used to
produce the EPR `paradox' \cite{epr},
where the `\mmt' on one part of a \sy fixes the state of another,
remote, part of the same \syn.

While the physical and philosophical discussion continues almost
unabated, over the years there have been what might be called
`practical' applications of the ``collapse" and the ``paradox". In
1968 Lipkin \cite{lipkin} proposed using it to study properties of
\k decays, and  in \b physics \cite{bigi}, where one
studies the decay  $\Upsilon(4s)\to \bn\bn$, it has been used to
study CP violation \cite{texp}. One uses the decay of one member of
the pair into a given flavor state to determine that the other
member is in the opposite flavor state, at the same time.
Furthermore, we recently explained
\cite{obs}  how the ``collapse''
can be used to circumvent spatial resolution difficulties connected
with the relatively  large size of the beam crossing region at the 
$e^+e^- $  colliders where the $\Upsilon(4s)$ is produced.

 While in the applications to  states of definite flavor one might
dismiss the results as consequences of  simple flavor conservation 
in strong interactions, one also has, as explained for example in
ref\;\cite{obs}, applications to other, more non-trivial, 
states of the two-state \syn.
In this paper we would like to study such further, more subtle
application of
the \cown, and to present a systematic formalism allowing a general
treatment of the \k\d or \b decays in the \sys  $\phi(1020)\to
\kn\kn$, $\psi(3770)\to \dn\dn$, or  $\Upsilon(4s)\to \bn\bn$. In
particular we will note applications to  CP
violation--- particularly {\it direct } \cvn.

   The concept of the \cow undoubtedly brings a number of
conceptual and pyschological difficulties with it, and we believe
these can be lifted by using the amplitude and not the \wf as the
fundamental quantity \cite{coll}. However, for the present purposes
it appears  convenient and  more  familiar to work in the \wf
approach, and in the following we will take the `collapse' quite
literally.  In section\,\ref{test} we mention an experimental test
of our interpretation of the  `collapse'. Finally,  it is possible
that the principle can be applied in other fields of
physics as well, but we shall not go into this here.

\section{The Principle}
Briefly, our idea is that the \cow of one part of a \sys can be
used to put the other part in a \qml state which reflects features
of the interactions involved in the 'collapse'. In this way small
effects, like
\cv in the `collapse' amplitudes, can be put into direct evidence.

  Let a
\sy consist of two coherent  parts, as in the  decay $
\phi(1020)\to \kn\kn.$
 A `measurement' on one part (we
will call this the `away side') fixes --`collapses'--the \wf of
that
part.  Often the other part (called the `near side')-- will be
connected to the first part by some symmetry or conservation
principle which correlates\footnote{In  quanto-babble these
correlations
are often called `entanglement'.} the \wfs of the two parts. The
`collapse' on the `away side' thus  determines or partially
determines the \wf of the `near side'. Hence by observing the
further behavior 
of the `near, uncollapsed, side' one can obtain information on the
interactions inducing  the `collapse'. We shall pursue the idea
that  this information can
include even small effects, like those due to \dcvn.

\section{The $\phi(1020)$,
$\psi(3770)$,  $\Upsilon(4s)$ \sys}

Because there is a high degree of symmetry and simplicity in the
p-wave decays of the $\phi(1020)$,
$\psi(3770)$, and $\Upsilon(4s)$ to a \pl anti-\pl pair, where each
is a two-state \syn, namely
$\kn,\dn$ or $\bn$ respectively, these are interesting and
experimentally accessible cases. In addition
to the interest of  the \qml principles, it can also offer a way to
determine some of the paramters of the  $\kn,\dn$ or $\bn$
\sysn. Indeed, the first suggestion of this
general kind, by Lipkin \cite{lipkin} concerned \cv 
in \k  decay.

The  $\kn,\dn$ or $\bn$  we
deal with is a
 two-state \syn, represented by a two-dimensional linear vector
space. This can be conveniently visualized as a kind of `spin 1/2',
an analogy we
shall use in the following.
 These \pls will in turn decay into certain final
states, the decay ``channels". We shall refer to the first \pl to
decay as the \as, and the undecayed  one, the one we may
follow further, the  \ns. 

We will utilize two major points concerning the two-state \k \d or
\b:

$1)$
 The decay into a
specific channel, like $\pi^+\pi^-$ or $\pi^o\pi^o$ for the \k, is
tantamount to a
`measurement' and fixes the state of the originating \sy (the 
$\kn $) in its
two-dimensional space; the `spin' is fixed in some definite
`direction'. Of course, all \wfs  can have an arbitrary overall
phase factor.

$2)$ Bose-Einstein statistics in the form of an overall symmetry of
the \wf applies to the identical
meson pair.  For the $\phi(1020), \psi(3770),$ or
$\Upsilon(4s)$ one has $l=1$ decays, so that the spatial \wf is
antisymmetric. Therefore the internal \wf of
the identical mesons must also be antisymmetric.  
This in turn implies that the two-state vectors of the mesons are
orthogonal. In section\,\ref{sym} we briefly examine the opposite
case, where the internal \wf is symmetric.

Another way of expressing 2) is to say that
given a certain
channel on one side, the other side (at the {\it same time}) must
be precisely  that state which {\it cannot} decay into the given
channel. Otherwise there would be a violation of Bose-Einstein
statistics----identical \sys in a p-wave \cite{lipkin}.
Note no further symmetries like CP are involved in this statement.

This statement 2)  describes the correlation mentioned above which
serves to fix
the `near side' once the `away side' has been `collapsed'. In other
applications the type of correlation can of course be different, as
with an  s-wave pair (section\,\ref{sym}). 

\section{Eigenstate for a Decay}\label{egn}
In appreciating the first statement  1) it is
important to recognize, 
as has been stressed in refs.\cite{obs},\,\cite{alv}, that a decay
channel, call it $a$,   defines a certain, unique, state of 
the two-state  \k \d or \b. 
For a two-state \sy there are
 two amplitudes for the  decay to  channel $a$.  In,
say, the flavor
basis  for the \k \syn, there is one amplitude
for the $\kn$  decay and one
for the $\kbn$ decay,  call
these amplitudes   $\alpha$ and $\alpha'$. Now with one channel and
two states, it
is always possible to
find a state which does {\it not } decay into the given channel:
namely the state $\alpha'\ket{\kn}-\alpha\ket{\kbn}$. This
state
has the decay amplitude $\sim \alpha \alpha'-\alpha'\alpha=0$ and
evidently does {\it not}  go into the channel $a$. On
the other hand, the state orthogonal to this no-decay state, namely
$\alpha^*\ket{\kn}+\alpha'^*\ket{\kbn}$, has the decay
amplitude
$\sim \vert \alpha \vert^2+\vert \alpha' \vert^2 $ and {\it does}
decay into the channel
$a$. Hence given a sample of events
with `away side'  $a$, the state vector of the `away side' is
uniquely
determined  (up to the overall phase
factor). We shall assume
that the fact that a decay has occured means the \sy was in that
state, with no component of the other, not-allowed-to-decay,
state. In section\,\ref{test} we propose a test of this assumption.

 At the same time, this also fixes the
orthogonal state or `\pl',  of the two-state \sy on the \ns;
it is the one which
does {\it not}  go into the channel $a$.

\section{Direct \cv}
Of course in the presence of some exactly conserved quantum number
such as CP,
 channels like $K^o \to \pi^+\pi^-$  and $K^o \to \pi^o\pi^o$ can
define the {\it same} state of the parent,
a state of definite CP, the CP=+1 state called
$\ket{K_1}=\frac{1}{\surd2}(\ket{\kn}+\ket{\kbn})$, and thus
determining the
orthogonal state to be
$\ket{K_2}=\frac{1}{\surd2}(\ket{\kn}-\ket{\kbn})$ on the `near
side'.

 However, if CP
is not exactly conserved,  we may expect
that the `away side' state vector
so determined is not precisely a state of pure CP, nor is it
necessarily the same for different channels  like $\pi^+\pi^-$ or
$\pi^o\pi^o$ . It is these possible small
differences we would like to examine for the study of \dcvn.

    One may thus anticipate a number of  experimental
consequences of \dcv on the \as \, which can be studied  in the
behavior of
the `near side'. These  will be discussed sytematically below, but
we mention two simple ones which come quickly to mind
 
 Two-Channel Difference-- Consider two different `away side'
channels,  of the same CP like $\pi^+\pi^-$
and $\pi^o\pi^o$ for the \k. If CP were perfectly conserved in the
respective `collapses', then both channels would define exactly the
same `near side' state at $t=0$, namely $K_2$. But with \dcv
in the `collapse', the \ns\, may be different for the two
channels. Hence {\it any}
differences at all in the evolution  of the `near side'  for the
two data
samples $\pi^+\pi^-$ or
$\pi^o\pi^o$ implies there was a CP violation on the `away side'.
(section\,\ref{2chan}).

Manifest CP impurities-- Decays on the `near side' can show a
manifest \pl--anti\pl asymmetry when the \as\, amplitudes have
\dcvn. This is
particularly simple at $t=0$, where  $t=0$  is the time of the
`collapse' of the `away side'(\eq{las}). 

 We stress that these CP violations  necessarily represent {\it
direct} \cvn, since only
direct decay
amplitudes, with no time evolution and thus effects of the
mass matrix, are involved.

\section{Paramaterization}

To proceed, we define 
a parameter $\zeta$ characterizing the CP violation in the `away
side' decay amplitude. Continuing to use
the \k \sy  to exemplify the ideas, we use the basis 
of CP eigenstates  $K_1,K_2$ and consider decays to
states  of 
definite CP. By  ``states  of 
definite CP" we mean those states, like $\pi^+\pi^-$ or $(J/\psi)
K_s$ which would be CP eigenstates in the CP conserving linit. For
explicitness we shall usually refer to $K_1,K_2$,  but the
discussion
  applies to  \d or \b equally well, using  $D_1,D_2$
or $B_1,B_2$ states.

  The \dcv will be manifested as an amplitude from the `wrong' CP
state into the decay channel in question. We thus define
\beql{zta}
\zeta = \frac{\alpha'^*}{\alpha^*}=\frac{(CP\,violating
\;amplitude)^*}
{(CP\,conserving\;amplitude)^* }
\eeql
 With a CP=+1 decay channel the numerator is thus the complex
conjugate of the  $K_2$ decay 
amplitude and the
denominator is the complex  conjugate of the $K_1$ amplitude. With
a CP=-1 decay
channel the
identifications are reversed. The connection of this
characterization of \dcv with the traditional notation is discussed
in the Appendix.

\subsection{States}

 Thus  the normalized
states which can decay into  a channel characterized by
$\zeta$ are for  CP=+1 channels,
$\ket{K_\zeta}=\frac{1}{\sqrt{1+|\zeta|^2}}(
\ket{K_1}+\zeta\ket{K_2})$, and  for  CP=-1, 
$\ket{K_\zeta}=\frac{1}{\sqrt{1+|\zeta|^2}}(\zeta\ket{K_1}+
\ket{K_2}).$
While in the $\zeta=0$ limit these states are of course orthogonal,
 in general there is no particular statement about the
orthogonality or not
of these two states. In any case the $\zeta$'s depend on the
decay channel in question.

In Table\,\ref{states} we show these states, together with their
orthogonal states. Since we will sometimes need the states in the
flavor basis we show this also.
It will be seen that the orthogonal states 
$\ket{K_{\zeta_\perp}}$ are just those that do not decay into the
given channel.

\begin{table}
\begin{center}
\begin{tabular}{|l|l|l|l|}
\hline
 Decay &&CP\,basis&Flavor\,basis\\
\hline
CP=+1&&&\\
\hline
&$\ket{K_\zeta}$&$\frac{1}{\sqrt{1+|\zeta|^2}}\biggl(
\ket{K_1}+\zeta\ket{K_2}\biggr)$&$\frac{1}{\sqrt{2(1+|\zeta|^2)}}
\biggl((1+\zeta)
\ket{\kn}+(1-\zeta)\ket{\kbn}\biggr)$\\
\hline
&$\ket{K_{\zeta_\perp}}$&$\frac{1}{\sqrt{1+|\zeta|^2}}\biggl(
-\zeta^*\ket{K_1}+\ket{K_2}\biggr)$&$\frac{1}{\sqrt{2(1+|\zeta|^2)
}}\biggl((1-\zeta^*)
\ket{\kn}-(1+\zeta^*)\ket{\kbn}\biggr)$\\
\hline
CP=-1&&&\\
\hline
&$\ket{K_\zeta}$&$\frac{1}{\sqrt{1+|\zeta|^2}}\biggl(
\zeta\ket{K_1}+\ket{K_2}\biggr)$&$\frac{1}{\sqrt{2(1+|\zeta|^2)}}
\biggl((1+\zeta)
\ket{\kn}-(1-\zeta)\ket{\kbn}\biggr)$\\
\hline
&$\ket{K_{\zeta_\perp}}$& $\frac{1}{\sqrt{1+|\zeta|^2}}\biggl(
\ket{K_1}-\zeta^*\ket{K_2}\biggr)$&$\frac{1}{\sqrt{2(1+|\zeta|^2)}}
\biggl((1-\zeta^*)
\ket{\kn}+(1+\zeta^*)\ket{\kbn}\biggr)$\\
\hline
\end{tabular}

\end{center}
\caption{States of the two-state \sy determined by a decay channel,
 where the channel has CP plus or minus, and \dcv
parameter
$\zeta$ (\eq{zta}). The
state is called $\ket{K_\zeta}$ and is shown in both the CP basis
and in the flavor basis. Also shown is the orthogonal state
$\ket{K_{\zeta_\perp}}$.
For the \d or \b cases, replace $K_1$ by $D_1$ or $B_1$, and \k
by \d or \b. }
\label{states}
\end{table} 
\subsection{ Probabilities}
To consider the further development of the \ns\, after the
`collapse', the evolution of the states  is
most conveniently expressed in a \dm formalism,  
as was used in ref\,\cite{obs}.  The
probability for a  state of the two-state \sy described by a \dm 
$\rho(1)$ to evolve
into one given by  $\rho(2)$ is
\beql{evl}
Prob(1 \to 2)= Tr[\rho(2){\cal M}(t)\rho(1){\cal M}^\dagger(t)]\,,
\eeql
where $\cal M$ is  the 2 x 2 time evolution operator, given by the
exponential of the mass matrix $M$ , ${\cal M}=e^{-iMt}$. $ M$ is
in general not hermitian and its eigenstates are not necessarily
orthogonal.
\eq{evl} gives the probability to obtain the state 2 at time $t$,
having had  the state 1 at $t=0$. It applies to  the one-body 
problem of the evolution of a single two-state meson; the
 application to our two-body problem comes in through the
assignment of the
states `1' and `2' as in \eq{evla}.
 
\subsection{ Density matrices}

To use \eq{evl} it is necessary to  have the \dm associated with
the states.
These may be obtained from 
$\rho=\ket{K}\bra{K}$ using
Table\,\ref{states} and relations like
$\ket{K_1}\bra{K_1}=\hf(1+\sgx)$ or
$\ket{K_1}\bra{K_2}=\hf(\sgz+i\sgy)$. 
We use standard identifications of the pauli matrices as in
ref\,\cite{obs}) where $\sgz$ is the flavor operator:
\,$\sgz\ket{\kn}=+\ket{\kn},\sgz\ket{\kbn}=-\ket{\kbn}$. In  Table
\ref{statesa} we show the \dms for the states of
Table\,\ref{states}.

\begin{table}
\begin{center}
\begin{tabular}{|l|l|l|l|l|}
\hline
 Channel &Density\,Matrix\,\,$\rho$
&$d_1=\frac{1}{1+|\zeta|^2}\times$
&$d_2=\frac{1}{1+|\zeta|^2}\times$&$d_3=\frac{1}{1+|\zeta|^2}
\times$\\
\hline
CP=+1&&\\
\hline
&$\rho(\zeta)=\ket{K_\zeta}\bra{K_\zeta}$ &$
(1-|\zeta|^2)$&
$-i(\zeta-\zeta^*)$&
$(\zeta+\zeta^*)$\\
\hline
&$\rho(\zeta_\perp)=\ket{K_{\zeta_\perp}}\bra{K_{\zeta_\perp}}$&$
-(1-|\zeta|^2)$&
$i(\zeta-\zeta^*)$&
$-(\zeta+\zeta^*)$\\
\hline
CP=-1&&\\
\hline
&$\rho(\zeta)=\ket{K_\zeta}\bra{K_\zeta}$&$
-(1-|\zeta|^2)$&
$i(\zeta-\zeta^*)$&
$(\zeta+\zeta^*)$\\
\hline
&$\rho(\zeta_\perp)=\ket{K_{\zeta_\perp}}\bra{K_{\zeta_\perp}}$&$
(1-|\zeta|^2)$&
$-i(\zeta-\zeta^*)$&
$-(\zeta+\zeta^*)$ \\
\hline
\end{tabular}

\end{center}
\caption{Density matrices for CP eigenstate decay channels with
\dcv given by the
parameter $\zeta$ (\eq{zta}),  together with those for the
orthogonal state. The $d$ are to be used in the representation 
\eq{rhoform}:  $\rho(d)=\hf(1+{\bf d\cdot\sigma})$.  }
\label{statesa}
\end{table}

The \dms for the orthogonal  states ``$\perp$"  give the `initial
states' on the
\ns \, produced by the `collapses'
on the \as. These are pure states (unless different `away sides'
with different $\zeta$'s
are averaged together). For $\zeta=0$ one sees that these $\rho$
reduce to the projection operators for the $CP=\mp 1$ states,
namely $\hf(1\mp\sgx)$.

 Writing the $\rho$ in the representation
\beql{rhoform}
 \rho(d)=\hf(1+{\bf d\cdot\sigma})\,
\eeql
we show the values of $d$ in Table \ref{statesa}.
One has ${\bf d^2}=1$ so that 
$\rho^2=\rho$ as needed for a pure-state. One further notes
that $\rho(\zeta)$ and $\rho(\zeta_\perp)$ are obtained from one
another by reversing the `spin' via  ${\bf \sigma \to -\sigma}$ or
$\bf d\to -d$, as
should be expected from the `spin up-spin down' analogy. 
Also the CP=+1 and CP=-1 cases are connnected by conjugating with 
$\sgz$, that is  $\rho \to \sgz\rho\sgz$, since $\sgz$ is the CP
`flip' operator.
\eq{rhoform} with $\bf d$ a unit vector gives the most general form
of
a \dm representing a pure state. This involves two free parameters,
 corresponding to the  real and imaginary parts of $\zeta$.

Calling $\bf \Delta$ the deviation of the $\bf d$ from the simple
values for CP tags so that  ${\bf d}=(d_1,d_2,d_3)=(\pm 1,0,0)+{\bf
\Delta}$  we
have
\beql{del}
{\bf \Delta} =\frac{1}{1+|\zeta|^2}(\mp 2|\zeta|^2,\,\pm 2{\cal
I}\{\zeta\},\,2 {\cal R}\{\zeta\} )\,,
\eeql
where ${\cal I}\{\zeta\}$ and $ {\cal R}\{\zeta\}$ refer to the
imaginary and real parts. For the orthogonal states one has ${\bf
d} \to -{\bf d}$ and so also a reversal of the sign of $\bf
\Delta$.   The normalization condition ${\bf
d^2}=1$ implies $\pm2\Delta_1=-\Delta^2$.

 The presence of a $\Delta_3$ via ${\cal R}\{\zeta\}$  implies a
flavor asymmetry which can be induced by the \cvn. This will be
manifested below
(section\,\ref{tzero} ) where we note how a CP tag on the \as\, can
lead to a flavor asymmetry at t=0 on the \ns, if ${\cal
R}\{\zeta\}\neq 0$.

\subsection{Flavor Tag}
The $\zeta$ notation, although   perfectly general, is oriented
towards   decays involving CP eigenstates with a small \dcv.
 However there is also the  frequently used lepton tag for  flavor
eigenstates such as
\k or \kb,
given by decays of the type $\kn \to l^+...$, where the sign of the
lepton implies the sign of the flavor. In this case we
will simply
 use  $\rho=\hf(1\pm \sigma_3)$, or  ${\bf d}=(0,0,\pm 1)$
(equivalent to
$\zeta=1$,  purely real). This is permissable since in the Standard
Model flavor change for the  flavor eigenstates requires higher
order weak
interactions, negligible on the order of the effects we discuss
here. This assumption has the consequence that with the lepton tag
we may reverse the sign of $\bf d$ by reversing the sign of the
lepton, a procedure that will prove useful below in contructing
various asymmetries. For  the CP tag, on  the other hand, there is
in general no obvious way of accomplishing an experimental reversal
of the sign 
of $\bf d$, once we allow for nonzero $\bf \Delta$ (see
sections\,\ref{tt} or \ref{ccpt}).

\subsection{Completeness}
Finally we note the completeness relation
\beql{compl}
I=\ket{\zeta}\bra{\zeta}+\ket{\zeta_\perp}
\bra{\zeta_\perp}=\rho{(\zeta)}+\rho{(\zeta_\perp)}\,,
\eeql
which we will use below to exchange a state with its orthogonal
state.

\section{Tags and  Rates}
\eq{evl} is the most simple and transparent quantity theoretically.
However it is not what can be directly measured experimentally.
We suppose an experimental procedure as follows. Let there be a
sample where the first, or \as \, decay   from the $l=1$ boson pair
is to a channel `a'. The time of each decay establishes a $t=0$. 
Then in a time interval $dt$ around a  later time $t$,  we  count
the
number of events in the second or \ns\, decay into a channel `b'.
The number of these second decays is proportional to the time
interval $dt$, so what we obtain from experiment is a rate quantity
we can call $ Rate(b,a;t)$, for the rate to `b' given `a' at $t=0$.
This quantity differs from
\eq{evl} in two ways. First, `a' refers to the \as, unlike  `1'
which refered to the `\ns'. Secondly, dealing with a rate means we
must introduce the rate constant $\Gamma_b$ which gives the rate of
decay into the channel `b ' from the eigenstate of the two-state
\sy for `b' decay.

As explained above, `a' on the \as\, implies the orthogonal state
$a_\perp$ on the \ns, whose \dm is found by reversing the sign of
$\bf d$. The formula for the experimental quantity $ Rate(b,a;t)$
is thus
 \beql{evla}
Rate(b,a;t)=\Gamma_b \,Prob(a_\perp \to b) =\Gamma_b \,
Tr[\rho({\bf d_b}){\cal
M}(t)\rho({-\bf d_a}){\cal
M}^\dagger(t)]\,,
\eeql
The introduction of the partial rate $\Gamma$'s will in some cases
make the examination of simple predictions derived for the $Prob$
more complicated, but in certain ratios involving different
channels the $\Gamma$'s can be made to cancel. A general procedure
for finding the $\Gamma$'s  experimentally is explained in
section\,\ref{pr}.

\subsection{Time ordering of the `Collapse'} \label{ordr}
An interesting question arises if we consider the  two decays--
`\mmtn
s'--\,separated by a short time, short compared with the internal
time evolution as governed by $\cal M$.
 If a decay `b'  takes place shortly after the
`collapse' `a'  on the other side
such that time evolution through the mass matrix has essentially no
time to take effect, then we should expect to get a closely related
result with the reverse time ordering. That is, where the `b'
`collapse' occurs first. After all, we just have  two
decays  `a' and `b', essentially at the same time. In fact we find
that \eq{evl} is the same at $t=0$, regardless of which `collapse'
is first.  We  use
\eq{compl},
$I=\rho(a)+\rho(a_\perp)$  to exchange a state with its orthogonal 
 or $-\bf d$ state   
\beql{ord}
Tr[\rho(a_\perp)\rho(b)]=Tr[\bigr(I-
\rho(a)\bigl)\bigr(I-\rho(b_\perp\bigl)]=
Tr[\rho(b_\perp)\rho(a)]\,,
\eeql
using $Tr[I]=2$ and $Tr[\rho]=1$. 
Thus one obtains  the same for  \eq{evl}, regardless of the
time ordering, for `collapses' separated by a short time.   It is
perhaps interesting to
note that these arguments on are on the level of
amplitudes squared and that there  is the possibility of a phase
factor, presumably unobservable, on the amplitude level. \eq{ord}
as such, is simply a property  of the one-body \eq{evl}, and is not
a
prioi connected to the two-meson problem.
Below we shall further consider  the exchange of `a' and `b' when
the time interval is not small.

\subsubsection{ Determination of the Partial $\Gamma$}\label{pr}
The above point also has an operational interest. While the most
simple  theoretical quantity is the $Prob$ or trace
expression \eq{evl}, the  most direct
experimental quantity is \eq{evla},  $ Rate(b,a;t)$. 
 However, this raises the difficulty that one must know the
partial rate $\Gamma_b$ to obtain the simple $Prob$ quantities.

 But now \eq{ord} gives a method to
find the various partial $\Gamma$; or rather their ratios to a
given common one.
Let the rate be measured for `a' on the \as, and `b' on the \ns.
And  inversely let the rate be measured for `b' on the \as, and `a'
on the
\ns. The ratio between the two is $ Rate(b,a;t)/Rate(a,b;t)$. 
Letting $t \to 0$ and using \eq{ord} 
\beql{rat}
\frac{Rate(b,a;t)}{Rate(a,b;t)} \to
\frac{\Gamma_b}{\Gamma_a}\frac{Tr[\rho(a_\perp)\rho(b)]}{Tr[\rho(
b_\perp)\rho(a)]}=\frac{\Gamma_b}{\Gamma_a}~~~~~~~~~~~~~~~~~~~~~~
t\to 0\,.
\eeql
Thus one may find the ratios of various partial $\Gamma$s in terms
of directly measureable  quantities. One
begins with either channel `a' or channel `b' on the \as , and
finds the rate to the other channel on the \ns . Taking the ratio
and extrapolating to $t=0$, one finds the ratio of the partial
$\Gamma$'s. Doing this for all
or many channels, one
can obtain all or many of the partial $\Gamma$ in terms of one of
them, a `calibration' rate. We emphasize
that the statement on the equality of the traces for $t\to 0$ is
a consequence of the \qm alone and does not involve any assumption
about symmetry properties of the interactions. We stress that these
partial $\Gamma$'s do not have any label for an initial state,
since they refer to only one state, the eigenstate for the decay
channel.

\section{Evaluation of the Trace}
Given the \dms  and the partial $\Gamma$'s, the rates or $Prob$'s
for any
process of `collapse' and detection may be evaluated----at least
relative to one `calibration' rate.

\subsection{Expansion of ${\cal M}$}
  It is possible to proceed somewhat further in the evaluation by
using
a standard identity to  expand the ${\cal M}=e^{-iM t}$ matrix.
First we remove the scalar part $\sim I$ of the matrix
${ M}=m-i\hf \Gamma$. The real scalar part 
cancels, leaving the total decay rate
$\Gamma$ so that \eq{evl}
becomes
\beql{ev2}
Prob(1\to 2)= e^{-\Gamma t}\, Tr[\rho(2)e^{-i{\bf M\cdot
\sigma}\,t}
\rho(1)e^{i{\bf
M^*\cdot \sigma}\,t}]\,.
\eeql
 $\bf M$ is the complex three-vector in the expansion of the
traceless part of $\cal M$, namely ${\bf M}\cdot \sigma=({\bf
m}-i\hf{\bf
\Delta \Gamma})\cdot \sigma$:
\beql{mdef}
{\bf M}={\bf m}-i\hf{\bf
\Delta \Gamma}\,.
\eeql
We note $({\bf M}\cdot\sigma)^\dagger={\bf M^*} \cdot \sigma$.
Now using the identity $e^{-i{\bf b\cdot\sigma}}= cos\, b -i ({\bf
b\cdot\sigma})\frac{sin\,b}{b}$, one arrives at three terms with
different time dependences
\beql{ev3}
Prob(1\to 2)= e^{-\Gamma t} \biggl( A\, \bigl|cos( M\, t)\bigr|^2
+  \biggl(B\frac{sin(M \,t)}{M} cos(M^* t) +cc\biggr)
+C\,  \biggl|\frac{sin Mt}{M}\ \biggr|^2 \, \biggr)\,.
\eeql
The `cc' refers to the complex conjugate of the $B$ term. 
 $ M$ is the complex  quantity arising in $ M ^2
={\bf M}^2=(m_1-\hf i\Delta \Gamma_1)^2+(m_2-\hf
i\Delta\Gamma_2)^2+(m_3-\hf i\Delta\Gamma_3)^2$.
For  $M\approx M^*$, one may write $\frac{sin(M \,t) cos(M^* t)
}{M}\approx \frac{sin(2M \,t) }{2M}$.
 and  $M$ is one-half the real mass
difference. With $\Delta\Gamma \neq 0$ the trignometric functions
in \eq{ev3} are to be understood as the full functions of a complex
variable.   For small \dg, one has to lowest, linear, order $ M=
m-\hf i\frac{\bf
m\cdot \dgn}{m}$, so that ${\cal R}\{M \}\approx m$ and  ${\cal
I}\{ M
\}\approx -\hf \frac{\bf m\cdot\dgn}{m}$.   ($\cal R$ refers to
real part, $\cal I$ to imaginary
part.) When $\bf m$ and \dg are parallel vectors, as is true for
the \k \sy to ${\cal O }\sim 10^{-3}$, then $ M= m-\hf
i\Gamma$.

For the coefficients one finds
\begin{multline}\label{abc}
A=Tr[\rho(2)\rho(1)]=\hf(1+{\bf d_2\cdot
d_1})~~~~~~~~~~~~~~~~~~~~~~~~~~~~~~~~~~~~\\
B=-iTr[\rho(2){\bf M}\cdot
\sigma\rho(1)]~~~~~~~~~~~~~~~~~~~~~~~~~~~~~~~~~~~
~~~~~~~~~~~~~~~~~~~~~~~~~~~~\\
=\hf ({\bf d_1\times d_2})\cdot{\bf M}-i\hf({\bf
d_1+d_2})\cdot{\bf M} \\
~~~C=Tr[\rho(2){\bf M}\cdot \sigma\rho(1){\bf M^*}\cdot
\sigma]~~~~~~
~~~~~~~~~~~~~~~~~~~~~~~~~~~~~~~~~~~~~~~~~~~~
~~~~~~\\
~~~~~~~~~=\hf{\bf M}\cdot{\bf M^*}+\hf{(\bf d_1-d_2)\cdot m \times
\Delta
\Gamma}+~~~~~~~~~~~~~~~~~~~~~~~~~~~~~~~~~~~~~~~~~~~~~~~~~~~~~~~\\
{\cal R}\{ {\bf (d_2 \cdot M)( d_1\cdot M^*)\}-\hf (d_2\cdot d_1)(
M\cdot M^*) }~~~~~~~~~~~~~
\end{multline}
With \eq{ev3} and \eq{abc} one has  a full, systematic,
description of all time dependent phenomena, applying to any of the
\k, \d, or \b \sysn. These may then be used for $\phi(1020)$,
$\psi(3770)$, and $\Upsilon(4s)$ to obtain a $Rate(b,a;t)$ by
inserting 
\beql{ins}     
{\bf d_1}=-{\bf d_a}~~~~~~~~~~~~~~~~~~~~~~~~~~{\bf d_2}={\bf
d_b}\,,
\eeql
according to \eq{evla}.

\subsection{At $t=0$, the $A$ term}\label{tzero}
The \ns\, at $t=0$ may be viewed as the starting
state provided by the `preparation of the \wf'
 resulting from the `collapse' on the \as, and this is reflected in
in the $A$ term of \eq{ev3}, which 
at $t=0$ is  the only surviving  term. $A$ is the only coefficient
which doesn't involve  $\bf M$ and so reflects only the properties
of the $\bf d$'s. In particular with CP tags, $A$ is sensitive to
deviations from the simple ${\bf d}=(\pm 1,0,0)$, showing \dcv, as
will be discussed for some cases below.

 For both
\as\, and \ns\, channels 
 the same, one has ${\bf d_2\cdot d_1}=-1$ or  $A=0$, and  so a
complete vanishing at t=0, as
expected from the Bose-Einstein statistics.

\subsection{Manifest Flavor and CP Asymmetries}
 An interesting configuration for 
\dcv is a CP tag on the \as\, say
${\bf d_1}=( 1,0,0)+{\bf \Delta}$ and the lepton or flavor tag on
the \ns,
${\bf d_2}=(0,0,\pm 1)$. One has $A=\hf(1 \pm\Delta_3)$. This
offers a way to obtain the real part of
the \dcv parameter $\zeta$ by taking the difference of the rates
for reversed lepton signs. Taking the difference for the 
two flavor tags one has
$Prob(l^+)-Prob(l^-)=\frac{2}{1+|\zeta|^2}{\cal R }
\{\zeta\}$. This may be
interpreted as saying that the impure CP state on the \as\, has
`prepared', via $\Delta _3$, a state on the \ns\, which is not
completely
neutral in flavor. With the lepton tag for flavor and using
\eq{evla} and assuming that to sufficient accuracy one has
$\Gamma(l^+)=\Gamma(l^-)$ for the leptonic decays, for  `a'  a
$CP=+1$  tag with parameter $\zeta$:
\beql{las}
\frac{Rate(l^+,a;t)-Rate(l^-,a;t)}{Rate(l^+,a;t)+Rate(l^-,a;t)}=
\frac{2}{1+|\zeta|^2}{\cal R }
\{\zeta\}~~~~~~~~~~~~~~~~~t\to 0 \,.
\eeql

In analogy to this production of a flavor asymmetry produced by a
CP tag one may consider a CP asymmetry produced by a flavor tag.
With no \dcv the oppopsite side to a flavor tag should have one
half of each CP, as reflected by $A(CP=+1)=A(CP=-1)=\hf$ in
\eq{ev3} following from ${\bf d_1}\cdot{\bf d_2}=0$ when there is
no \dcvn.  However with nonzero $\zeta$ we have $A(CP=+1)=\hf+
\Delta_3$. For the difference of two CP channels x and y $A^x-
A^y=\Delta_3^x-\Delta_3^y=\frac{2{\cal
R}\{\zeta^x\}}{1+|\zeta^x|^2}-\frac{2{\cal
R}\{\zeta^y\}}{1+|\zeta^y|^2}$. The x and y channels my be of same
or opposite CP, in any case the nonvanishing of the $A$ difference
is indicative of \dcvn.

\section{Exchange of ``Collapse and Detection'' }\label{ex}
Above, in section\,\ref{ordr}, we considered the question of
exchanging
`collapse' and `\dtn' for $t\to 0$.  We can   now  examine the
question
at  finite times, with propagation
effects included, using the above general
 expressions \eq{ev3} and \eq{abc}.  Let there  first  be  a
tag with channel `a' and parameter $\bf d_a$ on the \as, and then
afterwards a decay into channel `b' on the \ns.  Is there some
relation between this and the reverse situation, where the tag is
`b'  and the later decay is channel `a'?

  In   the exchange of ``Collapse'' and ``Detection'' one  has the 
configurations for the \dm parameters  in  \eq{abc}    
\begin{multline}\label{af}
~~~~~~~~~~`a'~first:~~~  {\bf d_1=-d_a}~~~~~~~~~~~~{\bf d_2=d_b}\\
`b'~first:~~~  {\bf d_1=-d_b}~~~~~~~~~~~~{\bf
d_2=d_a}~~~~~~~~~~~~~~~~~~~~~~~~~~~~~~
\end{multline}
 Under exchange of the two  cases, $B$  in \eq{abc} 
changes sign while $A$ and $C$ do not.
At the same time, we note that the $B$ term in \eq{ev3}  is odd
under $t \to -t$ while the $A$ and $C$ terms are even.
 Hence one can obtain the 
$Tr$ expression 
 for `$a$ first' from that for  `$b$ first'
by  putting $t\to -t$.
\footnote{ The equivalence  of the $\bf d$ exchanges and $t \to -t$
can also be demonstrated formally by the kind of
manipulations one uses in connection with
the usual $ T$ operation 
\cite{sakurai}. However, 
 the result here is not a consequence of $ T$
invariance, and holds even if there is a $ T$ violating
interaction, such as an $M_2$ term.}

 Thus  one may say that the joint process ``
decay to $a$ together with decay to $b$'' is given by a single
function $f(t)$ via
$e^{-\Gamma t} f(t)$,
with $t$ or $-t$ inserted in $f$ according to whether `a' or `b' is
first. (The $t$ is the time interval between the
two decays, and is always positive.) 
 Since $f(t)$ has both odd and even parts there is no particular
relation between $f(t)$ and $f(-t)$, the important
point is that $f$ is smooth, that the odd part vanishes at at
$t=0$.

Taking the difference of the two configurations only the $B$ term
contributes
\begin{multline}\label{abfa}
Prob(a_\perp\to b;t)-Prob(b_\perp\to a;t)=2 \times e^{-\Gamma
t}\times(-1)\\
\hf \bigl( ({\bf d_a\times d_b})\cdot {\bf M}-i ({\bf
d_a-d_b})\cdot {\bf M}\bigr)\frac{sin(Mt)}{M}cos(M^*t) +cc \,.
\end{multline}

 The expression vanishes identically for $\bf d_a \equiv d_b$ since
the two processes then become the same: `a' on the \as\, and `a'
on the \ns.
By the same token, \eq{abfa} is therefore sensitive to
deviations from the equality of the $\bf d$'s. This is significant
for the detection of \dcv since if we consider two channels with
the same CP, they have the same $\bf d$ unless there is  \dcvn.

\subsection{Both `a' and `b' are  CP channels}
 With both `a' and `b'  channels of the same  CP and no \dcv one
would have $\bf  d_a=d_b$ and zero for \eq{abfa}. \eq{abfa} is thus
an experimental quantity representing \dcvn, independently of \cv
in mixing. Operationally
it is obtained by comparing rates for a sample of events with `a
first' with a sample for `b first', and then using \eq{evla} to
find the $ Prob(a_\perp\to b;t)$ and $Prob(b_\perp\to a;t)$. If the
two are not equal (at all times) then
there is \dcvn.

Let there be two channels of the same CP 
like  $\pi^+\pi^-$ or
$\pi^o\pi^o$ in the \k \syn. Consider the
 quantity $\Delta\bf d=d-d'=\Delta-\Delta'$ for the difference of
their $\bf
d$ parameters. This is only nonzero if there is \dcv and is related
to  the $\zeta$ as  in \eq{del}. In \eq{abfa} it
leads to
\beql{afbc}
 - e^{-\Gamma t}\\
 \biggl({\bf d\times \Delta d}\cdot {\bf
M}-i {\Delta \bf d}\cdot {\bf M}\biggl)\frac{sin(Mt)}{M}cos(M^*t) 
+cc \,.
\eeql

Hence by exchanging the order of two channels of the same CP and
taking the difference one
finds their relative \dcv $\Delta\bf d$. With no \dcv the result
should be zero.
In the approximation where \dcv or $\bf \Delta$ is small so  $\bf
d$ is approximately in the $`1'$ direction, one then has 
that ${\bf d}\times\Delta\bf{ d}$ is either in the `3'
or `2' direction. The `3' component would multiply a CPT
violating $M_3$ and should be very small. This leaves a $\Delta_3
M_2$ term, which after taking the cc will principally involve
 ${\cal R}\{\zeta\} $. Information on ${\cal I}\{\zeta\} $ can come
from the
second term in \eq{afbc}.

 If `a' and `b' are  channels of opposite  CP and the $\bf
\Delta$'s are not large, the second  term in \eq{abfa} will
dominate and is proportional to $M_1$.

\subsection{Both `a' and `b' are Flavor Channels,  CPT
test}\label{cptt}
In the Standard Model the lepton tag is expected to select a state
of definite flavor to high accuracy, so that ${\bf d}=(0,0,\pm 1)$.
The first term in \eq{abfa} vanishes and the second term is given
by  the CPT forbidden $M_3$.
Hence the non-vanishing of 
\beql{ll}
Rate(l^-,l^+;t)-Rate(l^+,l^-;t) \sim i
M_3\frac{sin(Mt)}{M}cos(M^*t)+cc \,,
\eeql
 indicates a CPT violation, essentially proportional to ${\cal I}\{
M_3\}$.
One has assumed 
$\Gamma_{l^+}=\Gamma_{l^-}$ and ${\bf d}=(0,0,\pm 1)$ for the
flavor tags. Since this test amounts to the difference between a
flavor and an anti-flavor process, it  corresponds to a known test
in \k physics \cite{kcpt}.

\subsection{One Flavor, One CP Channel, Identification of $\cal
I\{\zeta \}$}\label{oneone}
In this case \eq{abfa} can be non-zero  with no \dcvn. The first
term is proportional to the $ T$ violating $M_2$ and the second
to  the $ T$ conserving $M_1$. This first term is what
arises in simple CP and $ T$ tests in the \b \sy \cite{obs}.

 Deviations from these simple values  are of interest in obtaining
$\Delta_2$, or  ${\cal I}\{\zeta \}$.
In \eq{abfa} the $({\bf d_a\times d_b})$ term, as said, is in the
`2' direction without \dcv and so is proportional  to $M_2$, as in
the simple simple CP, $ T$ tests discussed in ref\,\cite{obs}.
However with \dcv $({\bf d_a\times d_b})$ will contain a term $\pm
\Delta_2$ in the `1 direction, proportional
 to $M_1$. The resulting  deviation from the simple, purely mixing
induced, predictions for these tests thus provides a measurement of
${\cal I}\{\zeta \}$.

\section{Same \as \, and \ns} \label{na}
In the previous section \ref{ex} we considered the exchange of
`collapse' and 'detection' channels. Continuing with some general
features, we note
that a particularly simple situation arises when both are
the same channel, the `same on both sides' configuration. Then we
have  ${\bf d}_1=-{\bf d}_2$  and so 
\eq{abc} becomes
\begin{multline}\label{abc2}
A=0\\
B=0~~~~~~~~~~~~~~~~~~~~~~~~~~~~~~~~~~~~~~~~~~~~~~~~~~~~~~~~~~~~~~
~~~~~~~~~~~~~~~~~~~~~~~~~~~\\
C=
{\bf M}\cdot{\bf M^*}- {\bf d\cdot m \times \Delta
\Gamma}-{\bf (d \cdot M)( d\cdot M^*)}~~~~~~~~~~~~~~~
~~~~~~~~~~~~~~~~~~~~~\,,
\end{multline}
where ${\bf d}={\bf d_1}$ gives the \dm for the \as\, or
equivalently the final
state on the \ns.

The vanishing of A and B in \eq{abc2} is the generalization of the
statement that
the same state cannot appear on both sides at t=0. In
particular this leads to the feature that the rate is given by C
alone, so that the small $t$ behavior for the rate is necessarily
 $\sim t^2$ for any choice of channel.

  These formulas are completely
general, and in particular independent of the presence of \dcv with
a nonzero
$\zeta$ or not. Thus {\it any} data set with the same \as\, and \ns
\,will have the same $e^{-\Gamma t} \biggl|\frac{sin(
M\,t)}{M}\biggr|^2 $ behavior. This parallel behavior  applies to
all three \sys ($\phi, \psi, \Upsilon$) and to all
channels, independent of CP or CPT questions, and
 provides a test of the \qml procedure.

 Since $M$ and $\Gamma$ are general
properties of the \syn, the only difference between various
channels used will be in their $C$ parameters and of course
their partial decay rates, which can be found via the method of
section\,\ref{pr}. If we compare two channels 
$x$ and $y$,  both used in the `same on both sides' configuration,
one has
\beql{ss}
\frac{Rate(x,x;t)}{Rate(y,y;t)}=\frac{\Gamma_x}{\Gamma_y}\frac{C(
x)}{C(y)} \,,
\eeql
 constant in time. 

\subsection{Lepton tag}
Continuing with the  same \ns and \as\, configurations, one expects
to high accuracy that a lepton tag as in $\kn\to
l^{\pm}+...$ specifies the flavor, and  so  ${\bf d}$ to be
$(0,0,\pm 1)$.
Therefore  one has from \eq{abc2}  that $C$ 
becomes
\beql{cl}
C(l^\pm)=
{\bf M}\cdot{\bf M^*}-\pm ( {\bf m \times \Delta
\Gamma})_3- |M_3|^2=|M_1|^2+ |M_2|^2-\pm ( {\bf m \times \Delta
\Gamma})_3
\eeql
where a possible CPT violating $M_3$ term cancels out.
Thus in the difference between lepton signs only the cross term
survives and one will have
\begin{multline}\label{sl}
\frac{Rate(l^+,l^+;t)-Rate(l^-,l^-;t)}{Rate(l^+,l^+;t)+Rate(l^-
,l^-;t)}=\frac{\Gamma_{l^+}C(l^+)-\Gamma_{l^-}C(l^-
)}{\Gamma_{l^+}C(l^+)+\Gamma_{l^-}C(l^-)}\\=\frac{C(l^+)-C(l^-
)}{C(l^+)+C(l^-)} =\frac{( {\bf m \times \Delta
\Gamma})_3}{|M_1|^2+ |M_2|^2}~~~~~~~~~\,,
\end{multline}
where we use  $\Gamma_{l^+}=\Gamma_{l^-}$.   Thus in the `same on
both sides'  configurations this
kind of ratio can be used to obtain information on 
\dg. Or more precisely on to the extent that \dg and $\bf m$ are
not parallel, i.e do not commute. For the \b, \eq{sl} should be
very small,  in $\psi(3770)\to\dn\dn$  it may help to obtain
information on the \d mass matrix.

 A point to note here is that the numerator in \eq{sl} is bounded.
Since     ${(\bf m \times \Delta\Gamma})_3\leq m\Delta\Gamma$,
there is a limit to the ratio, which presumably can be determined
in a complete experimental analysis. Also ${(\bf m \times
\Delta\Gamma})$ has only the `3' component when CPT is good, since
the both vectors lie in the `1-2'plane.

In the presence of \dcvn, the analogous statement to \eq{sl} is not
rigorously
possible with  CP tags. This is because  if a given decay channel
corresponds to a certain $\bf d$, there is not necessarily another
channel that corresponds to $-\bf d$. This is further discussed in 
section\,\ref{ccpt}.

\subsection{\dg $\approx 0$}\label{dg0}
 When we can put \dg $\approx 0$, as for the \b \syn , C in
\eq{abc2}
further simplifies:
\beql{abc3}
C=  m^2  - {\bf (d \cdot
m)}^2~~~~~~~~~~~~~~~~~~~~~~\dgn\approx 0
\eeql
 For CP tags with no \dcv, $\bf d$ is $(\pm 1,0,0)$  
  and so $C$  is constant from channel to channel. A
variation is then indicative of \dcvn.  For \dcv  negligble,
and assuming good CPT  one will thus 
have $C= m_2^2$, the square of the CP,  T violating
mass term. This affords a direct measurement of $m_2$ to the
accuracy allowed by the assumptions $\dgn\approx 0$ and $\Delta
\approx 0$.
Deviations from this simple value at more than the $10^{-3}$ level
\cite{acc}  should give information on the
other, \dcv components of the $\bf d$. This should be one of the
simplest ways to observe \dcv in the \b \syn. 

\subsection{Same CP on Both Sides}

A variation on the  `same on both sides' configuration is 
 `same CP on both sides'. If one takes channels of the
same CP, like $\pi^+\pi^-$ and $\pi^o\pi^o$ for the \k, in the
absence of \dcv they both define the same state of the \k. Then one
has the `same on both sides' configurations, as  discussed in the
earlier parts of this section. In particular one will have $A=0$
and the corresponding vanishing of the rate towards $t=0$.
On the other hand, with \dcv the two are no longer identical \sysn,
and the rate need not vanish  towards $t=0$. In particular one has,
with parameters   $\bf d$ and $\bf d'$,
\beql{samecp}
A=\hf(1-{\bf d}\cdot{\bf d'})=(1/4) (\bf 
\Delta -\Delta')^2 
\eeql
where we used the normalization relation mentioned after \eq{del}.

This gives, via \eq{ev3}, a non-vanishing rate at t=0, This was
essentially Lipkin's proposal \cite{lipkin}. 
 The value at $t=0$ measures
the "non-identity" of the two channels. It is amusing that is
possible, by means of this process, to give a quantitative measure 
of the extent to which two different states are `identical'. 

\section{$B^o\, System$} \label{gz}
Another general simplification ensues when one may
take $\Delta
\Gamma\approx 0$.
 In the $\Upsilon(4s)\to \bn \bn$ \sy 
this is believed to be true to about the $10^{-3}$ level
\cite{acc}. Setting \dg=0 one has $\bf M$ real, $\bf M= m$ and $\hf
\Delta
M=\hf\surd m^2$ real numbers and so 
  the  ordinary trignometric functions in
\eq{ev3}.In particular one now has for the $B$ term $(B+B^*)
cos(mt)sin(mt)/m
=(B+B^*) sin(2mt)/2m$ and 
the coefficients 
\begin{multline}\label{abc1}
A=\hf(1+{\bf d_1\cdot d_2})\\
~~~~B+B^*= ({\bf d_1\times d_2})\cdot {\bf m}
~~~~~~~~~~~~~~~~~~~~~~~~~~~~~~~~~~~~~~~~~~~~~~~~~~~~~~~~~~~~~~~~~
~~~~~~\\
C= \hf m^2 (1-{\bf d_1\cdot d_2})  
+  {\bf (d_1 \cdot m)( d_2\cdot
m)}~~~~~~~~~~~~~~~~~~~~~~~~~~~~~~~~~~~~
\end{multline}
The general configurations examined above may be discussed here
with these
simplified forms. The same \as\, and \ns\, was discussed in
section\,\ref{dg0}.

The exchange of \ns\, and \as\, of section\,\ref{ex}
now becomes 
\begin{multline}\label{abfb}
Prob(a_\perp\to b;t)-Prob(b_\perp\to a;t)=2 \times e^{-\Gamma
t}\times(-1)\\
({\bf d_a\times d_b})\cdot {\bf m}\,\frac{sin(2
m \,t)}{2m}\,.
\end{multline}
With two flavor tags, ${\bf d}=(0,0,\pm 1)$, this quantity is zero.
If it were nonzero at more than the  $10^{-3}$ level , this could
indicate a larger than expected
value for \dg.

With two CP tags, ${\bf d}=(\pm 1,0,0)+\bf \Delta $, the leading
terms
would be linear in the  $\bf \Delta $.  The `2'
component of $\bf \Delta_a-\Delta_b$ multiplies the CPT violating
$M_3$, so that largest term should be
$ (\Delta_a-\Delta_b)_3 M_2$. Since the CP, or T violating $M_2$ is
not
small \cite{obs}, this offers a way to obtain the channel
differences
${\cal R} (\zeta_a-\zeta_b)$. There is also a $ (\Delta_a\times
\Delta_b)$ term, presumably small.

The case of one flavor tag and one CP tag will be discussed in
section\,\ref{tt}.

\section{ $ T$ tests }\label{tt}
 A simple and intuitive $ T$ test consists in comparing a
certain process "forwards" and "backwards". In ref\,\cite{obs},
where \dcv was neglected, in section 5, "T asymmetry", for example,
one compared $\bn \to B_2$
with $B_2 \to \bn$. In the language we use here, both \pls were on
the \ns. With neglect of \dcv one could hope to identify the $B_2$
in the final state by a channel thought to be predominately CP=-1
and to produce it in the initial state by a channel thought to be
predominately CP=+1 on the \as.

However with \dcv taken into account, as we do here, it is
not clear how to do this. Given a certain final channel on the \ns,
there is no certain means to make this also the initial state on
the \ns\, by some \as\, tag.

 However we can propose an analogous test here, with \dcv taken
fully into account. One uses  the exchange of \ns\, and \as\, as
examined in
section\,\ref{ex}.
The quantity  \eq{abfa} resembles that used in the discussion of
the previous $ T$ tests. However the $\bf d$ now include a
possible \dcv via $\Delta$, \eq{del}.

  Applying \eq{abfa} to the $\Upsilon(4s)\to \bn\bn$ case, with
neglect of \dg,  one has that the effect is simply given by 
 $({\bf d_a\times d_b})\cdot {\bf M}$, \eq{abfb} . With two flavor
tags
the $\bf d$ are both in the $\pm 3$ direction and one obtains zero
\cite{wolf}. We thus consider the flavor tag-CP tag
configuration,
 In the limit of no \dcv for this configuration one has  ${\bf
d_a\times d_b}$ in the 
`2' direction, and the effect is proportional to $m_2$, the CP, T
violating term in the mass matrix. One thus obtains essentially the
same result as in ref\,\cite{obs}. Here, including \dcv with a
nonzero $\bf \Delta$ for the CP tag, there is also a term $\Delta_2
m_1$. 

 Thus in the $\Upsilon(4s)\to \bn\bn$ \sy an asymmtery between the
cases `flavor tag first, CP tag second' and vice-versa requires
either an $m_2$ or an imaginary \dcv, to the level that \dg can be
neglected. 
A detailed measurement of the effect would be of interest since
finding the
$\Delta_2 m_1$ contribution would provide a way of obtaining $\cal
I\{ \zeta\}$, while most of our simple effects are proportional to
$\cal R\{ \zeta\}$.

\section{Two-Channel Differences}\label{2chan}
As mentioned
earlier, one of the simplest manifestations of \dcv is in the
comparison of two 
\as\,  channels of the same CP.
 {\it Any} difference at all on the \ns\, indicates \dcv.

In our formulas, these differences result from having different
$\Delta$'s for
the $\bf d$ of the two \as\, tags. Since the final state on the
\ns\, is the same in the two cases, $\Gamma_b$ will drop out in a
ratio such as
\beql{asy}
\frac{Rate(b,a;t)-
Rate(b,a';t)}{Rate(b,a;t)+
Rate(b,a';t)}=\frac{Prob(b,a;t)-
Prob(b,a';t)}{Prob(b,a;t)+Prob(b,a';t)}\,,
\eeql 
where  we call the different \as\, channels of the same CP $a$ and
$a'$.

 Any nonzero value of the experimental quantity on the lhs
indicates \dcv. The connnection to $\bf d_a$ and $\bf d_{a'}$ and
thus the $\zeta$ may be found by writing  ${\bf d_a}=(0,0,\pm
1)+{\bf \Delta_a}$ and ${\bf d_{a'}}=(0,0,\pm 1)+{\bf
\Delta_{a'}}$, and finding the differences in the coefficients
A,B,C when  (the negative of) these are inserted in \eq{abc} for
$\bf d_1$ and one sets $\bf d_2=\bf d_b $.

\section{CPT Tests with CP tags}\label{ccpt}
Good CPT requires that the `3' component of $\bf M$ be zero,
$M_3=0$, which also implies that ${\bf m \times \Delta\Gamma}$ has
no `1,2' components. In searching for such a forbidden component,
in section\, \ref{cptt} we used the property that  with a lepton
tag  it is possible to reverse the sign of $\bf d$ by reversing the
sign of the lepton.

 With CP tags this would be also possible to the extent that one
negelcts \dcv for the decay channels, that is one puts $\bf
\Delta=0$. Then one may reverse $\bf d$ by simply choosing a
channel of the opposite CP. For example in \eq{abc2}, one has the
term
${\bf d\cdot m \times \Delta\Gamma}$, which is odd under a reversal
of $\bf d$. Let  the $Prob$'s and so the $C$'s in the `same on both
sides' configurations be determined  for two channels of opposite
CP, which we designate with `+' and `-'. Then
\beql{sscp}
\frac{Prob(+)-Prob(-)}{Prob(+)+Prob(-)}=\frac{C(+)-C(-)}{C(+)+C(-)}
=\frac{\bf (m \times
\Delta\Gamma)_1}{|M_2|^2+|M_3|^2}~~~~~~~~~~~~~~~~~{\bf
\Delta}\approx 0
\eeql
provides an expression sensitive to the CPT forbidden
$\bf (m \times \Delta\Gamma)_1$.

This test is only good only to the level that the assumption of no
\dcvn, that is ${\bf \Delta}=0$, holds. However it may be possible 
with extensive data and analysis of the type described in this
paper that the values of ${\bf \Delta}$ can be established in some
channels to good accuracy. Then inserting these in \eq{abc2}, the
necessary corrections to \eq{sscp}  could be found.

 Similar arguments apply to other instances where the forbidden
$M_3$ or $\bf (m \times \Delta\Gamma)_{1.2}$ appear. In fact, it is
conceivable that the appearance of nontrivial $\bf \Delta$ makes
new kinds of tests possible. For instance in the above example, if
a large  $ \Delta_2$
is established for some channel, then there would be a sensitivity
to the CPT violating  ${\bf (m \times \Delta\Gamma)}_{2}$.

\section{ The $l=0$ System}\label{sym}
The $l=0$ \syn, where the internal \wf will be symmetric,  instead
of antisymmetric,  is of conceptual interest, although it is not
clear if there is a useful experimental situation of this type.  
 It would result from the decay of a spin-zero
resonance  into a pair of \k \d or \b \plsn. We consider it briefly
as an illustration of  the general scope of the
problem.

 The $l=1$ case with the antisymmetric internal state, as we have
discussed at great length, is actually the simpler case. For the
following reason. There is only one antisymmetric  state of two
two-state \sys. In the \k  notation this is
$\frac{1}{\surd2}(\ket{K}\ket{\bar K}-\ket{\bar K}\ket{K})$.
In the spin-1/2  analogy   this is the singlet
state, with total `angular momentum' zero,  of the two `spins'. Any
evolution will simply return the same state since the evolution --
rotation of the `spins'--will respect the symmetry and there is
just this one state. On the same grounds, this state expanded in
terms of the state for a given tag $\ket{K_\zeta}$, always  has the
same form,
namely $\frac{1}{\surd2}(\ket{K_\zeta}\ket{K_{\zeta_\perp}}-
\ket{K_{\zeta_\perp}}\ket{K_\zeta})$. This observation is the basis
of our repeated use of the fact that the state recoiling  against
a given tag is always the state orthogonal to that tag.

 With a symmetric internal state, the situation is different. There
are three possible symmetric states,   in the angular momentum 
analogy S=1 with $S_3=\pm 1,0 $. Namely  $ (\ket{K}\ket{K},   
\ket{\bar K}\ket{\bar K}$, and $\frac{1}{\surd2}(\ket{K}\ket{\bar
K}+\ket{\bar K}\ket{K})$.  True, due to flavor conservation in
strong interactions the pair will be `born' in this last, flavor
neutral, state. In the spin analogy this  is an eigenstate of $S_3$
with eigenvalue zero. Indeed, in the limit of no \cv the effect of
the mass matrix is to induce a rotation around the `1' axis and
this state is invariant--so that in this limit one would always
retain the same state.
 However, with \cv the rotation is not just around the `1' axis
 and other components will develop with time. Furthermore,
the form of an expansion in terms of  $\ket{K_\zeta}$ states will
not be invariant, as it is in the antisymmetric case. Hence in
general the state recoiling against a given tag will be some
 time-dependent combination of the same state and its orthogonal
state. And not simply the same state, as one might have naively
supposed.

 A complete analysis of this problem would involve starting from a
$t=0$ defined by the decay of the parent resonance and determining
the subsequent evolution of the two-body \sy due to the mass
matrices. While for the $l=1$  cases  it is
sufficient to count the time from the decay of one member of the
pair, here the analysis would have to start with the decay of the
parent
resonance, and  it is  not clear if this time  can be determined
sufficiently well experimentally.  If a suitable resonance is found
this problem might be worth further discussion.

This example shows that in  general the identification
of a recoiling state in terms of an \as\,\,tag is not always
elementary and must be examined on a case-by-case basis.

\section{Fundamentals of \qm}

Our arguments are within
the usual canon of \qmn, except, perhaps, that here the
"measurement" is via a spontaneous decay process and
does not involve any obvious active disturbance by an `observer'.
Nevertheless, the arguments
  use the ``collapse'' --or its equivalent in terms of
ampitudes \cite{coll})-- in very fine, perhaps unprecedented, 
detail. The confirmation and consistency of our predictions, here
as well as those  in  ref\,\cite{obs}, would
provide an impressive verification  of \qml principles.

\subsection{Test of Consistency of the  `Collapse'
Treatment}\label{test}
 Our most important assumption is that
 a decay fixes
 the \as\,  of the \k \d or \b pair.  Out of the
continuous range of possibilities available for the state vector of
the two-state \syn,
the \as \,state is fixed to be one only, the eigenstate  for the
decay channel. It is here that there appears to be an abrupt change
from `potentialities' to `certainty'.

While it is difficult to imagine another way of doing this, one
might entertain the thought that the \as\, decay does not imply the
corresponding eigenstate (see section\,\ref{egn}) with exactly
probability 1; perhaps the factor could be channel dependent or
dependent on the \ns\, state.

 It is thus interesting  that there is a test of the
consistency of
this assumption.
 This arises from the observation that there is more
than one way to arrive at the ratio of two partial rates.
 As discussed
in sect.\ref{pr}, one can find $\Gamma_a/\Gamma_b$ by taking the
ratio of rates for `a' on the \as\, and `b' on the \ns\, to that
for the inverted situation, and letting $t\to 0$. Let us call the
ratio of  partial rates determined by this experimental
procedure $\{\frac{\Gamma_a}{\Gamma_b}\}$:
\beql{rr}
\biggl\{\frac{\Gamma_a}{\Gamma_b}\biggr\}\equiv
\frac{Rate(b,a;t)}{Rate(a,b;t)}\biggr|_{t\to 0}
\eeql

  Now  it should be possible to find the
same ratio in a roundabout way via another pair of processes,
as in
\beql{pairs}
\biggl\{\frac{\Gamma_a}{\Gamma_b}\biggr\}=\biggl\{\frac{\Gamma_a}
{\Gamma_c}\biggr\}\times
\biggl\{\frac{\Gamma_c}
{\Gamma_b}\biggr\}
\eeql
The  procedures implied on the right are different from that on
the left and involve different channels, and it is perhaps
conceivable that experiment leads to different
numbers for the two
sides of  \eq{pairs}. Thus \eq{pairs} presents an experimental test
of our method  and in particular of our treatment of the \cown.
Naturally if tests of the type \eq{pairs} are experimentally
consistent it does not necessarily imply the veracity of the
method. But a clear breakdown of \eq{pairs}, or its
generalizations,  would be very interesting
and necessitate a  great rethinking  of the problem.

Finally, concerning the ultimate meaning of the \cown, we would
like
to take this opportunity to reiterate our view   \cite{coll}  that
the \cow is  a convenient fiction that arises due to an
unnecessary
reification of the wavefunction. The need for it goes away in
an amplitude
approach to \qmn, where there is nothing to `collapse' in the first
place. However, as one sees in the present
application, it is a very 
 convenient fiction, often allowing a quick and easy insight in
seemingly complicated situations.

\section{Conclusions}
 We have pursued the idea that the observation of a decay of a two-
state \sy like \k \d or \b amounts to a "measurement" that fixes
its internal state. In the $l=1$ decays
$\phi(1020)\to \kn\kn$,
$\psi(3770)\to \dn\dn$, or  $\Upsilon(4s)\to \bn\bn$,  Bose-
Einstein statistics then determines the recoil 
at the time of the decay to be  the  orthogonal state. This
implies that
the recoiling state will contain information on \dcv in the first
or \as\, decay. A general parameterization for such effects is
given and then applied to the further evolution of the recoiling
state, called the \ns. The
parameterization 
gives a clear separation of \dcv effects and mixing-induced \cvn.

The result of the analysis is a rich phenomenlogy and
some configurations 
of special interest are identified, particularly for studying
\dcvn. These include
exchange of  \ns\, and \as, `same on both sides',  same 
CP on both sides, and comparison
of two \as\, tags of the same CP.

Examination of a hypothetical analogous case  with $l=0$ shows that
in general the identification of the state of a recoil by this
method is not always elementary.

The method involves a quite detailed use of the \cown, and
experimental  results on the many predictions would provide tests
of the underlying ideas.
 A consistency test for the  treatment of the \cown, which
can be carried out by a certain determination of partial decay
rates, is suggested.

\section{Appendix}
Traditional notation for \dcv has been rather heterogeneous
and differs from case to case, often combining the effects of \dcv
and mixing
\cv in one parameter.
Our notation, using  the $\zeta$ parameter as expressed through
$\bf \Delta$, \eq{del},  offers a systematic, uniform description,
which applies to all cases. With $\bf \Delta$ or equivalently
$\zeta$ zero,
there is no \dcvn. Mixing-induced \cv is given
by $M_2$ the coefficient of $\sgy$ in the mass matrix, ${\bf
M}\cdot \sigma=({\bf
m}-i\hf{\bf
\Delta \Gamma})\cdot \sigma$:
The essential difference in the notations is that ours, adapted to
the ``collapse'' viewpoint, refers to the CP eigenstates, while the
conventional notation refers to the quasi-stationary states such as
$K_S,K_L$. Since the quasi-stationary states  are defined by the
mass matrix, there is inevitably a combination of effects when
these states are used.

 Thus while $\zeta$ resembles the conventional $\eta$ in
its definition, see \cite{eta}, $\eta$
contains effects both due to mixing in the mass matrix (the
$\epsilon$ parameter) and \dcv (the $\epsilon'$ parameter). The
connection   between an $\eta$ and a $\zeta$ 
can be obtained by expanding the states like $K_S,K_L$ in terms of
the $p,q$ parameters. This leads to
\beql{conn}
\eta=
\frac{\frac{1}{\surd 2}(p-q)+\frac{1}{\surd 2} (p+q)\zeta^*}{
\frac{1}{\surd 2}(p+q)+ \frac{1}{\surd 2}(p-q)\zeta^*} \,.
\eeql 
for a CP=+1 channel. For a CP=-1 channel, the sign of q is
reversed.

One notes that in the no-\cv-in-mixing limit, where $p=q$,   one
has
$\eta=\zeta^*$, as should be expected in accordance with the role
of $\zeta$ as  purely characterizing \dcvn. Otherwise $\eta$
combines both mixing and \dcv effects.

 In the limit of small $(p-q)$ and small $\zeta$, as in the \k
\syn,
one has, for example for the $\pi^o\pi^o$ or the $\pi^+\pi^-$ 
channels 
\beql{conna}
\eta_{\pi^o\pi^o}=
\frac{p-q}{p+q} + \zeta_{\pi^o\pi^o}^*
~~~~~~~~~~~~~~~~~~~~\eta_{\pi^+\pi^-}=
\frac{p-q}{p+q} + \zeta_{\pi^+\pi^-}^* \,.
\eeql

With  the traditional notation where $\eta_{\pi^o\pi^o}=\epsilon 
-2 \epsilon'$ and $\eta_{\pi^+\pi^-}=\epsilon + \epsilon'$, one has
$\epsilon'= \frac{1}{3}(\eta_{\pi^+\pi^-}-
\eta_{\pi^o\pi^o})=\frac{1}{3}(\zeta_{\pi^+\pi^-}^*-
\zeta_{\pi^o\pi^o}^*)$. This shows that a nonzero
$\epsilon'$ requires not only \dcv but also a difference between
the two channels involved.   With strong \cv in the
mass matrix, as in the \b  \syn,
 the full relation \eq{conn}
must be used to connect $\zeta$ and $\eta$.

 For another example, now in the \b system, ref\,\cite{bigi},
 in Eq.10.38, following \cite{cs}, discuss a flavor
tag on the \as, followed by detection of a CP state on the \ns. The
rate at $t=0$ (their $\Delta t=0$) is given by an $|A(f)|^2$ where
`f' refers to the final CP tag. In our notation  this is given by
the $A$ coefficient in  \eq{abc1}, $A=\hf(1 +\Delta_3)=\hf +{\cal
R}\{\zeta\}$. Thus  if the final tag involves a \dcv it has been
implicitly incorporated in $A(f)$, while in our notation it is
explicitly exhibited.  As explained in
section\,\ref{tzero}, this \dcvn, proportional to ${\cal
R}\{\zeta\}$ can be observed by reversing the sign of the lepton
tag or by comparing two different tags of the same CP.

\end{document}